\documentstyle[preprint,aps,epsf]{revtex}

\newcommand{\bb}{\begin{equation}}
\newcommand{\ee}{\end{equation}}
\newcommand{\bqn}{\begin{eqnarray}}
\newcommand{\eqn}{\end{eqnarray}}
\newcommand{\pp}{\partial }

\begin{document}
\draft
\preprint{\begin{tabular}{l}
ULB-TH-98/10
\end{tabular}}

\title{ Comments on Chiral $p$-Forms\footnote{To appear 
in the Proceedings of the Conference
``Quantum Gravity in the Southern Cone II", Bariloche (Argentina),
January 6-10, 1998.}}
  \author{Xavier Bekaert$^a$ and Marc Henneaux$^{a,b}$}
\address{$^a$Physique Th\'eorique et Math\'ematique, Universit\'e Libre de
  Bruxelles, \\
  Campus Plaine C.P. 231, B--1050 Bruxelles, Belgium,\\
  $^b$Centro de Estudios Cient\'\i ficos de Santiago,
  Casilla 16443, Santiago 9, Chile.
  }
\maketitle

\begin{abstract}
Two issues regarding chiral $p$-forms
are addressed. First, we investigate the topological conditions
on spacetime under which the action for a non-chiral
$p$-form can be split as the sum of the actions for two chiral
$p$-forms, one of each chirality.  When these conditions are
not met, we exhibit explicitly the extra topological
degrees of freedom and their couplings to the chiral modes.
Second, we study the problem of constructing Lorentz-invariant
self-couplings of a chiral $p$-form in the light of the Dirac-Schwinger
condition on the energy-momentum tensor commutation relations.
We show how the Perry-Schwarz condition follows 
from the Dirac-Schwinger criterion and point out that consistency
of the gravitational coupling is automatic.
\end{abstract}

\vfill


\tightenlines

\section{Introduction}
\setcounter{equation}{0}
\setcounter{theorem}{0}

[The talk by M.H. at the Bariloche meeting was devoted to
the results obtained in \cite{DGHT1,DGHT2,DHS} on the flip of
sign in the quantization condition for $k$-brane dyons
($k$ odd) in $2k+4$ dimensions. To avoid repetition with
what can be found in the literature, the present contribution
to the proceedings does not reproduce the actual content of the talk
but deals with the Lagrangian formulation of chiral $p$-forms.
We refer the reader interested in
the quantization condition  for $k$-brane dyons to
\cite{DGHT1,DGHT2,DHS,Stelle} for a detailed discussion.  See also
\cite{BJ,Ferr,Iengo} for related information].

\vspace{.5cm}

Chiral $p$-forms, i.e., $p$-forms, the field
strength of which is self-dual, can exist in
$(2p+2)$-Minkowski spacetime for any
even $p$.  They are notoriously known to suffer from one
major difficulty:  even though their equations
of motion are manifestly Lorentz-invariant, there
is no simple (e.g. quadratic in the free case),
manifestly Lorentz-invariant Lagrangian
that leads to these equations of motion \cite{MS}. 

Although there is no simple manifestly Lorentz-invariant Lagrangian,
there is a simple non-manifestly Lorentz-invariant
Lagrangian which has been given in \cite{HT1,HT2}, and
which generalizes the Lagrangian of \cite{FJ} for
chiral bosons.  This Lagrangian is linear in the first order 
time derivatives of the spatial components of the $p$-form potential
and reads, in the $2$-form case that we shall consider for definiteness
\begin{equation}
S[A_{ij}] = \int dx^0 d^5x B^{ij} \partial_0 A_{ij}
- \int dx^0 H \; \; \; (i,j, \dots = 1, \dots, 5)
\label{action0}
\end{equation}
with 
\begin{equation}
B^{ij} = \frac{1}{3!} \epsilon^{ijklm} F_{klm}, \; \; F_{\mu \nu \lambda}
= \partial_\mu A_{\nu \lambda} - \partial_{\nu} A_{\mu \lambda}
- \partial_\lambda A_{\nu \mu}
\label{defB}
\end{equation}
and
\begin{equation}
H = \int d^5x ( N {\cal H} + N^k {\cal H}_k).
\end{equation}
Here, $N$ and $N^k$ are the standard lapse and shift \cite{MTW}.
The magnetic field $B^{ij}$ is a spatial tensor
density of weight one.
We are considering from the outset the theory in a gravitational
background as in \cite{HT1,HT2}.
In the absence of self-interactions, the energy density ${\cal H}$ is given by 
\begin{equation}
{\cal H} = \frac{1}{\sqrt{g}} B^{ij} B_{ij}
\label{Hbis}
\end{equation}
where the spatial indices are lowered and raised with the spatial
metric and its five-dimensional inverse, while $g$ is
the determinant of $g_{ij}$.  The energy density generates displacements
normal to the slices of constant $x^0$.  The momentum density ${\cal H}_k$,
on the other hand, is purely kinematical and generates tangent
displacements.  It is explicitly given by
\bb
{\cal H}_k = \frac{1}{2} \epsilon_{ijmnk}B^{ij}B^{mn}. 
\ee
In order to write the action (\ref{action0}), it is necessary to assume that
spacetime has the product form ``time $\times$ space". 
This will be done throughout in the sequel.

The action (\ref{action0}) is manifestly invariant under
the gauge transformations
\bb
\delta_{\Lambda} A_{ij} = \partial_i \Lambda_j - \partial_j \Lambda_i 
\label{gaugetransf}
\ee
since $B^{ij}$ is gauge-invariant and identically
transverse ($\pp_i B^{ij} \equiv 0$)\footnote{Since
$A_{0i}$ does not occur in the action -- even if one replaces
$\partial_0 A_{ij}$ by $\partial_0 A_{ij} -
\partial_i A_{0j} - \partial_j A_{i0}$ (it drops out because
$B^{ij}$ is transverse) --, the action is of course invariant under
arbitrary shifts of $A_{0i}$.  
It is also invariant under arbitrary
shifts of any other field that does not
appear in the action.}.
In flat space, it is also invariant under Lorentz transformations,
but these do not take the usual form \cite{HT1,HT2}.

Knowing the action for a chiral form in a gravitational
background, one can compute the gravitational anomaly
by usual quantum field theoretical methods 
\cite{BvN} and compare the result with calculations based
on the non-chiral action supplemented by an appropriate
projection \cite{AGW}.  As shown in \cite{BvN}, there is agreement.

The first part of our paper is motivated by this result and
aims at understanding better the
relationship between the non-chiral action and 
the chiral ones. 
We show  that 
when the spatial sections have vanishing
Betti numbers $b_2$ and $b_3$, 
the action for a non-chiral
form is just the sum of the actions for two
uncoupled chiral forms of opposite chiralities.
Thus the path integral for a non-chiral $2$-form
supplemented by a projection to one chiral sector,
trivially reduces to the path-integral for the
corresponding chiral modes.
This is no
longer true for more general topologies.  The non-chiral action
and the sum of the
chiral actions agree on the local degrees of
freedom, but treat differently
the harmonic components of the $2$-form.  However, one
can easily keep track of the topological
``zero mode" difference.
This is explicitly done in section \ref{comp},
after we have reviewed the necessary background on the
dynamics of chiral $p$-forms. 
The importance of global features
when dealing with chiral forms has been pointed out and stressed
in \cite{Witten1} where the problem of
modular invariance has been addressed.  Recent developments
relevant to the six-torus case are given
in \cite{Dolan-Nappi}.

In an interesting series of papers \cite{PST}, a manifestly
covariant formulation of chiral $p$-forms has been developed.
This formulation is characterized
by the presence of an extra field and an extra gauge invariance.
This extra field occurs  non-polynomially in
the action, even for free chiral $2$-forms.  The manifestly
covariant formulation has proved useful for many conceptual
developments.
It has been shown to be equivalent to the non-manifestly covariant
treatment of \cite{HT1} in Minkowski space \cite{PST2}. To the extent
that the analysis of \cite{PST} strongly relies on 
the Poincar\'e lemma, 
it is expected to share
also similar global features.

The second question analysed in this paper
is that of Lorentz-invariant self-couplings
(as well as consistent self-couplings in an external gravitational background)
for chiral $p$-forms.  In view of its relevance to the $M$-theory
five-brane, this question has received a lot of attention, both
at the level of the equations of motion \cite{EOM} and at
the level of the action \cite{PerryS,S,Sall,PST5,Ceder}.
We show that this question can be handled by
means of the Dirac-Schwinger condition on the commutation
relations of the components of the energy-momentum tensor
\cite{Dirac,Schwinger}.  
This condition leads directly to
the differential equation obtained in 
\cite{PerryS} and implies automatically
consistency of the gravitational coupling.
So, once Lorentz-invariant 
self-interacting chiral $p$-form theories have been found, there
is no extra work to be carried out to couple them to gravity.  The 
Dirac-Schwinger criterion, which appears to be
quite powerful in the present context, has been used recently 
in \cite{DeserS}, in the investigation of Lorentz-invariance
of manifestly duality-invariant theories in
the other even ($0$ mod $4$) spacetime  dimensions.

\section{Dynamics of chiral $2$-form}
\setcounter{equation}{0}
\setcounter{theorem}{0}

As stated above, we assume that spacetime takes
the product form $T \times \Sigma$ where $T$ is the manifold
of the time variable (usually a line). Furthermore,
we also assume that the spatial sections $\Sigma$ are either
homeomorphic to $R^5$ (in which case the theory must be supplemented by
fall-off conditions at infinity that insure the vanishing of the relevant
surface terms), or compact. Of course, a spatial coordinate could
equivalently play the r\^ole of the time variable, as in \cite{PerryS}.

We define the exterior form
$B$ to be the (time-dependent) spatial $2$-form
with components $B_{ij}/\sqrt{g}$.
The equations of motion that follow
from the action are \cite{HT1,HT2}
\bb
d[N(E-B)] = 0
\label{EB}
\ee
where $E$ is the electric spatial $2$-form defined through
\bb
E_{ij} \equiv 
 \frac{\dot{A}_{ij} - N^s F_{sij}}{N}
\label{defE}
\ee
and where $d$ is the spatial exterior derivative operator.
In the case where the second Betti number $b_2 $ 
of the spatial sections vanishes, 
this equation implies
$N(E-B)=dm$, where $m$ is an arbitrary spatial $1$-form.  
To bring this equation to a more familiar form, one
sets $m_i = A_{0i}$.  The equations
of motion read then
\bb
F= \! ^{*} F
\ee
where $F_{0ij} = \dot{A}_{ij} - \pp_i A_{0j} + \pp_j
A_{0i}$.  This is the standard self-duality condition.
Alternatively, one may use the gauge freedom to
set $m=0$, which yields the self-duality condition
in the temporal gauge.

To deal with the case where $b_2$ is not zero,
one uses the Hodge decomposition of exterior forms on the spatial
sections \cite{math}.
Any form - and in particular, any $2$-form - can
be written as the sum of an exact form, a co-exact form
and a harmonic form,
\bb
A = d\rho + \delta \phi + \sum_A \lambda_A(t) \omega^A.
\ee
Here, the codifferential $\delta$ acting on a $p$-form
is equal to
$\delta = (-1)^{5p} \; * d *$, while
$\rho$ (respectively  $\phi$) is a spatial
$1$-form (respectively, spatial $3$-form) and
$\{\omega^A\}$ is, on each spatial slice,  a 
basis of harmonic (= closed and co-closed) $2$-forms.  These satisfy
$\partial_i (\omega^{Aij} \sqrt{g}) = 0$, $\partial_{[i}
\omega^A_{jk]}= 0$ and are normalized so that $\int d^5x \omega^{Aij}
\omega^B_{ij} \sqrt{g} = \delta^{AB}$ for each $t$.
The harmonic forms are in finite number and thus, the
harmonic component of $A$ describes a finite number of
global ``zero modes".  In the simple case where
$b_2=0$, there are no zero modes.  In the case where
$b_2 \not= 0$, $A$ may have a non-trivial harmonic
part.
The equation of motion (\ref{EB}) implies in that case
\bb
N(E-B)= \sum_A k_A(t) \omega^A + dm.
\label{E-B}
\ee
Again, one can absorb the exact part of the right-hand side
of (\ref{E-B}) in a redefinition
of $E$ (or set it equal to zero by a gauge transformation),
but there is an additional piece which is not determined, namely,
the harmonic part.  However, this harmonic part turns
out to be pure gauge, because the action (\ref{action0})
for a chiral $2$-form has more gauge invariances than
expressed by (\ref{gaugetransf}).  It is actually invariant under addition
to $A$ of an arbitrary closed (and not necessarily exact) $2$-form,
\bb
\delta_{\lambda,\epsilon} A = d \lambda + \epsilon_A \omega^A.
\label{transf}
\ee
This follows because $B$ is co-exact (and not just co-closed),
and invariant under (\ref{transf})\footnote{If the spatial metric 
depends on $t$, the $2$-form $\omega^A_{ij}$ will be also time-dependent.
The time derivatives $\dot{\omega}^A_{ij}$ are clearly closed so that
$\int B^{ij} \dot{\omega}^A_{ij} d^5x = 0$.}.
One can thus gauge away the harmonic part of $N(E-B)$ and get
again the self-duality condition. Therefore, the action
(\ref{action0}) leads to the correct self-duality condition
but is a theory in which the zero modes of $A$ are pure
gauge (no physical component along the harmonic forms).
A similar phenomenon was described in \cite{HT2} for
chiral bosons on a circle.  
To summarize: for a chiral $2$-form, both the exact and the
harmonic components (i.e., the closed part of $A$) are
pure gauge  and it is the co-exact part only that
contains the physical degrees of freedom.

For an anti-chiral $2$-form, the action is\begin{equation}
S[A_{ij}] = - \int dx^0 d^5x  B^{ij} \partial_0 A_{ij}
- \int dx^0 H
\label{action1}
\end{equation}
with $H = \int d^5x (N {\cal H}' + N^k {\cal H}'_k)$.  
The energy density ${\cal H}'$ is the same as for a 
chiral form, but the momentum density ${\cal H}'_k$
differs in the sign.  The analysis proceeds exactly
as above and one finds this time the anti-chiral condition
\bb
E + B = 0.
\ee
An anti-chiral $2$-form described by the action (\ref{action1}) has
no physical harmonic component.

For later purposes, we shall need the brackets of the
gauge-invariant magnetic fields $B^{ij}$.
The orthodox way to proceed is to define conjugate momenta
and follow the Dirac method for constrained systems \cite{DMethod}.  
The chirality
condition appears 
as a mixture of second class constraints
and of first class constraints, the first class part being
related to the gauge invariance of the theory \cite{HT2}.  One may
work out the Dirac bracket of the gauge-invariant
fields by using the Dirac formula, but 
one may shortcut the whole procedure
and directly read the brackets from the action (\ref{action0}),
which is already in first-order form.  Either way, one finds
as Dirac brackets
(we consider the chiral case for definiteness, the
anti-chiral one differs in the sign)
\begin{equation}
[B^{ij}({\bf x}), B^{mn}({\bf x'})] =
\frac{1}{4} \epsilon^{ijmnk} \delta,_k({\bf x} - {\bf x'}).
\label{brackets}
\end{equation}

We shall also need the brackets of the energy densities
${\cal H}({\bf x})$ at two different space points.
A direct calculation using only
the form of ${\cal H}$ and the brackets (\ref{brackets}) yields
\begin{equation}
[{\cal H}({\bf x}), {\cal H}({\bf x'})] =
({\cal H}^k({\bf x}) + {\cal H}^k({\bf x'})) \delta,_k({\bf x} - {\bf x'}).
\label{DS}
\end{equation}
The relation (\ref{DS}), derived first
on general grounds in \cite{Dirac,Schwinger}, is deeply connected to
Lorentz-invariance and gravitational coupling
and we shall return to it below.

\section{Zero modes of a non-chiral form}
\setcounter{equation}{0}
\setcounter{theorem}{0}

The action for a non-chiral $2$-form is
\bb
S[A_{\mu \nu}] = - \frac{1}{2 \cdot 3!} \int d^6x 
\sqrt{-\, ^{6}\! g} F^{\lambda \mu \nu}
F_{\lambda \mu \nu}.
\label{action2}
\ee
We keep the same notations for the $2$-form, even though
$A_{\mu \nu}^{here} \not= A_{\mu \nu}^{before}$ (see relationship
(\ref{change}) below between non-chiral and chiral $2$-forms).
It is invariant under the gauge transformations
\bb
\delta_{\Lambda} A_{\mu \nu} = \pp_\mu \Lambda_\nu -
\pp_\nu \Lambda_\mu
\label{gaugetransfbis}
\ee
which enable one to set $A_{0i}$ equal to zero.
The exact part of $A_{ij}$ can then be also gauged away at any
given time, {\em but the harmonic part cannot}.  Indeed, the
action (\ref{action2}) is {\em not} invariant under shifts
of $A_{ij}$ by an arbitrary closed form, only under shifts of $A_{ij}$ by
an arbitrary exact form\footnote{More precisely, the transformation
$\delta_\epsilon A_{ij} = \epsilon_A(t) \omega^A_{ij}$ of
the spatial components cannot be supplemented by a transformation
of $A_{0i}$ such that $\delta_\epsilon F_{0ij} = 0$ for
arbitrary $\epsilon$'s.  Indeed, this would require
$\dot{\epsilon}_A \omega^A_{ij} + \epsilon_A \dot{\omega}^A_{ij}
= \hbox{exact form}$, which  forces $\epsilon^A$ to
be a solution of the differential equation $\dot{\epsilon}_A +
t_A^B \epsilon_B = 0$, where $\dot{\omega}^A=
t^A_B \omega^B + d(\hbox{something})$, showing that $\epsilon^A$
cannot be an arbitrary function of time.   
The transformations with  
$\epsilon$ solution to this equation should be
regarded as rigid symmetries, not gauge symmetries.}.
Thus, the harmonic part of a non-chiral
$2$-form describes true physical degrees of freedom.

It is easy to see that on a flat background,
the harmonic part of $A_{ij}$
behaves like a free particle, i.e., grows linearly with
time,
\bb
A = d\rho + \delta \phi + \sum_A \lambda_A(t) \omega^A
\ee
with
\bb
\lambda_A(t) = C_A t + D_A
\ee
on-shell.  This is because the equation $\partial_\mu 
(\sqrt{- \, ^{6} \! g} F^{\mu \nu \sigma})
= 0$ implies $\frac{d^2{\lambda}_A}{dt^2} \omega^A = \delta(\hbox{something})
+ d(\hbox{something'})$
and thus $\frac{d^2{\lambda}_A}{dt^2}=0$.
Now, if the integration constant $C_A$ is different from zero,
the form $A_{\mu \nu}$ cannot be purely chiral or anti-chiral.  Indeed, if it
is chiral (say), then, the chirality condition implies
\bb
C_A \omega^A + d \dot{\rho} + \delta \dot{\phi} =
\delta (\hbox{something})
\ee
which leads to a contradiction unless $C_A = 0$.
Accordingly, if one decomposes the field strength into self-dual part
and anti-self-dual part, there is {\bf no} potential neither
for the self-dual part, nor for the anti-self-dual part
when $C_A \not= 0$, although there is a potential for the sum.

The situation is the same as for a chiral boson $\varphi$ on a circle.
The zero mode $\varphi_0 = at+b$ cannot be written as the
sum of single-valued left-movers and right-movers unless $a=0$,
even though the sum is single-valued ($at = (a/2)(t + \sigma)
+ (a/2) (t - \sigma)$ but $t+ \sigma$ or $t - \sigma$ are not single-valued).
Of course, the field strength $F_\mu = \partial_\mu \varphi$ is 
decomposable into well-defined self-dual and anti-self-dual parts, but
these do not derive from a single-valued potential.

We thus see that a non-chiral $2$-form contains additional global degrees
of freedom besides the local degrees of freedom described by the local
chiral actions.  It is the presence of the 
physical zero modes that is responsible 
for the fact that the sum of a chiral $2$-form and
an anti-chiral $2$-form is not a non-chiral $2$-form
on a topologically non-trivial background.

\section{Decomposition of non-chiral action}
\label{comp}
\setcounter{equation}{0}
\setcounter{theorem}{0}

In order to compare the action for a non-chiral $2$-form with
the sum of the chiral and anti-chiral actions given above, it
is convenient to rewrite the non-chiral action
in Hamiltonian form.  
To that end, one follows the Dirac method.  One finds
\bb
S[A_{ij}, A_{0i}, \pi^{ij}] = \int dx^0 d^5x [\pi^{ij} \dot{A}_{ij}
- \frac{N}{\sqrt{g}}( \pi^{ij} \pi_{ij} + \frac{1}{4} B^{ij} B_{ij})
- N^k \pi^{ij} F_{kij} - 2 A_{0i} \pi^{ij}_{,j} ]
\label{canonical}
\ee
where $\pi^{ij}$ is the momentum conjugate to $A_{ij}$.
The component $A_{0i}$ appears as a Lagrange
multiplier for Gauss'law constraint
\bb
\pp_i\pi^{ij} = 0.
\label{Gauss}
\ee

One can solve Gauss'law for $\pi^{ij}$ and eliminate the corresponding
multiplier from the action.  The general solution of (\ref{Gauss})
is
\bb
2 \pi^{ij} = \frac{1}{2}\epsilon^{ijklm} 
\partial_k Z_{lm} + \sqrt{2g} \mu_A \omega^{Aij} 
\label{Z}
\ee
While the $2$-form $A_{ij}$ is determined up to an exact form,
the $2$-form $Z_{ij}$ is determined up to a closed form,
\bb
\delta_{\Lambda',\chi} Z_{ij} = \partial_i \Lambda'_j
- \partial_j \Lambda'_i + \chi_A \omega^A_{ij}.
\ee
Using (\ref{Z}) and making the change of variables of
\cite{DGHT2},
\bb
Z_{ij} = \sqrt{2} (U_{ij} + V_{ij}), \; \;
A_{ij} = \sqrt{2} (U_{ij}  - V_{ij})
\label{change}
\ee
one finds, after straightforward algebra
\begin{eqnarray}
S[U_{ij}, V_{ij}, \mu_A] &=& S^{chiral}[U_{ij}]
+ S^{anti-chiral}[V_{ij}] \nonumber \\
&+&
\int d^6x \mu_A \sqrt{g} \omega^{Aij}(\dot{U}_{ij} - \dot{V}_{ij})
- \frac{1}{2}\int d^6x N \mu_A \omega^A_{ij}
\epsilon^{ijklm} \partial_k (U_{lm} + V_{lm})
\nonumber \\
&-& \frac{\sqrt{2}}{2} \int d^6x N^k \sqrt{g} \mu_A \omega^{Aij} F_{kij}
- \frac{1}{2}\int dt k^{AB} \mu_A \mu_B
\label{actionfinal}
\end{eqnarray}
with
\bb
k^{AB} = \int d^5x \sqrt{g} N \omega^A_{ij} \omega^{Bij}.
\ee
The action for the non-chiral form splits thus as the sum of 
two chiral actions, one
of each chirality, plus terms coupling the zero modes $\mu^A$ to
the chiral components.  The action is
invariant under the transformations
\begin{eqnarray}
\delta_{X,\xi} U_{ij} &=& \partial_{[i} X_{j]} + \xi_A(t) \omega^A_{ij}
\nonumber \\
\delta_{X,\xi} V_{ij} &=& \partial_{[i} Y_{j]} + \xi_A(t) \omega^A_{ij}
\nonumber \\
\delta_{X,\xi} \mu^A &=& 0
\end{eqnarray}
with {\em same} harmonic component $\xi_A(t)$ 
for $\delta_{X,\xi} U_{ij}$ and $\delta_{X,\xi} V_{ij}$.
Consequently, because of the zero mode coupling, 
the action (\ref{actionfinal})
has less gauge invariances than the sum of two chiral actions.
The zero mode of the difference $U_{ij} - V_{ij}$
is also gauge invariant.  One easily verifies that it is
in fact canonically conjugate to $\mu^A$.
For a flat metric, the couplings between the local degrees
of freedom and the zero modes simplify because the motion is an
isometry so that the time-derivative of a harmonic form is
harmonic.  One can
disantangle the zero modes from the co-exact
ones, but this will not be done here.

When $H^2_{DR} \not= 0$,
the physical Hilbert space for a non-chiral
two-form is bigger than the product of the Hilbert spaces for
a chiral two-form and an anti-chiral one.
One must also include the states associated with the harmonic modes,
\bb
{\cal H}^{non-chiral} = {\cal H}^{chiral}
\otimes {\cal H}^{anti-chiral} \otimes {\cal H}^0.
\ee
The truncation to the chiral sector is particularly simple
when there is no global, topological zero modes, since
it simply amounts then to dropping the uncoupled anti-chiral degrees
of freedom.  How to handle the global modes in the general
case depends on the context and will not be addressed here.

For issues that depend on the local (high-energy)
behaviour of the theory,
such as anomalies in local symmetries, the topological modes
should not be relevant.  
In the absence of such modes, the change of variables (\ref{change})
can be implemented easily in the path
integral and yields
\begin{eqnarray}
Z &=& \int DA D \pi  \exp{i(S[A, \pi])} 
\nonumber \\
&=& \int DU DV \exp{i(S^{chiral}[U] + S^{anti-chiral}[V])}  
\end{eqnarray}
where the measures $DA D \pi$ and $DU DV$ involve of course
the ghost modes and gauge
conditions.  Note that neither the change of variables
(\ref{change}) nor the parametrization (\ref{Z}) (when there is
no $\omega^A$)
involves the metric.
Projecting out to the chiral sector
by interting a delta-function $\delta(\pi^{ij} -  B^{ij})$ 
of the chirality condition
is equivalent to setting the
anti-chiral component $V_{ij}$ to zero, leaving one with the
path-integral for a chiral $2$-form.
Thus, implementing the chiral condition by a projection or
dealing with the non-manifestly invariant chiral action are clearly
equivalent in the absence of harmonic modes.

\section{Lorentz-invariant self-couplings and self-couplings
in a gravitational background}
\setcounter{equation}{0}
\setcounter{theorem}{0}

When one can use the tensor calculus, it is rather
easy to construct interactions that preserve Lorentz invariance.  These
interactions should also preserve the number of (possibly deformed)
gauge symmetries (if any), but this aspect is rather immediate for $p$-form
gauge symmetries -- although it is less obvious for the
extra gauge symmetry of \cite{PST}.

There is an alternative way to control Lorentz
invariance.  It is through the commutation relations
of the energy-momentum tensor components.  Because the
energy-momentum tensor is the source of the gravitational field, the
method gives at little extra price a complete grasp on the
gravitational interactions. 
As shown by Dirac and Schwinger \cite{Dirac,Schwinger}, a sufficient
condition for a manifestly rotation and translation
invariant theory (in space) to be 
also Lorentz-invariant is that its energy
density fulfills the commutation relations (\ref{DS}).
The condition is necessary when one turns to gravitation.
The method is more cumbersome than the tensor calculus when
one can use the tensor calculus, but has the advantage
of being still available even when manifestly invariant
methods do not exist.

In the Dirac-Schwinger approach,
the question is to find the most general ${\cal H}$ fulfilling
(\ref{DS}).  The energy-density ${\cal H}$ must
be a spatial scalar density
in order to fulfill the kinematical commutation relations
$[{\cal H}({\bf x}),{\cal H}_k({\bf x'})] \sim
{\cal H}({\bf x'})\delta,_k({\bf x} - {\bf x'})$ and depends
on $A_{ij}$ through  $B_{ij}$ in order to be
gauge-invariant.
In five dimensions, there are only two independent
invariants that can be
made out of $B_{ij}$,
\begin{equation}
y_1 = - \frac{1}{2g} B_{ij} B ^{ij}, \; \;
y_2 = \frac{1}{4g^2} B_{ij} B^{jk} B_{km} B^{mi},
\end{equation}
as can easily be seen by bringing $B_{ij}$ to canonical
form by a rotation (only $B_{12}$ and $B_{34}$ non zero; note
that in this local frame the only non-vanishing component of
${\cal H}^k$ is ${\cal H}^5$).
Set
\begin{equation}
{\cal H}= f(y_1,y_2)\sqrt{g}, \; f_1 = \partial_1 f, \;
f_2 = \partial_2 f.
\end{equation}

Then, a calculation following the standard pattern
and paralleling the free case calculation yields 
\begin{equation}
[{\cal H}({\bf x}), {\cal H}({\bf x'})] =
(\Lambda({\bf x}) {\cal H}^k({\bf x}) +
\Lambda({\bf x'}) {\cal H}^k({\bf x'})) \delta,_k({\bf x} - {\bf x'})
\end{equation}
with
\begin{equation}
4 \Lambda = f_1^2 + y_1 f_1 f_2 + (\frac{1}{2} y^2_1 - y_2) f_2^2
\end{equation}
Requiring that (\ref{DS}) be fulfilled gives
\begin{equation}
f_1^2 + y_1 f_1 f_2 + (\frac{1}{2} y^2_1 - y_2) f_2^2 = 4 
\end{equation}
which is precisely the equation (31) of Perry \& Schwarz with
$f$ replaced by $2f$.  The Dirac-Schwinger criterion yields thus directly the 
Perry-Schwarz equation, whose solutions are investigated in \cite{PerryS}.

In the flat space context ($g_{ij}= \delta_{ij}$, $N=1$, $N^k=0$), 
the equation (\ref{DS}) guarantees that
the interactions are Lorentz-invariant and no further work is
required \cite{Dirac,Schwinger}.  It also guarantees complete
consistency in a gravitational background because of
locality of ${\cal H}$ in the metric $g_{ij}$ \cite{Teitel,Teitel2}.

\section*{Acknowledgements} 
One of us (M.H.) is grateful to the organizers of the
meeting ``Quantum Gravity in the
Southern Cone II" for their invitation
and to the ``Laboratoire de Physique Th\'eorique
de l'Ecole Normale Sup\'erieure" for kind hospitality while
this work was carried out.
We thank Stanley Deser, Bernard Julia and Christiane
Schomblond for useful comments.  M.H. is
also grateful to Claudio Teitelboim for very
fruitful discussions in the early stages of this work.

\end{document}